\documentclass[twoside,onecolumn,11pt,a4paper]{book} 

\usepackage{anuthesis}
\usepackage{fancyhdr}
\usepackage{graphicx}
\usepackage{amsmath}
\usepackage{amsfonts}
\usepackage{amssymb}
\usepackage[sectionbib]{chapterbib}
\usepackage{appendix}
\usepackage{color}
\usepackage{cancel}
\usepackage{tensind}
\usepackage{feynmp-auto}
\usepackage{tensor}
\usepackage{dsfont} 
\usepackage{enumitem}
\usepackage{hyperref}
\usepackage{tikz}
\usetikzlibrary{decorations.markings}
\usepackage{subcaption}
\usepackage{physics}
\usepackage{algorithmicx}
\usepackage{amsthm}
\newcommand{\pau}[2]{\sigma^{#1}_{#2}}

\renewcommand\thesection{\arabic{section}}

\usepackage{cite}

\newcommand{\co}[1]{\left [ #1 \right ]}
\newcommand{\pa}[1]{\left ( #1 \right )}

\newcommand{\lla}[1]{\left \{ #1 \right \} }

\newcommand{\ord}[1]{\mathcal{O}(#1)}

\newtheorem{theorem}{Theorem}[section]

\newtheorem{Def}[theorem]{Definition}

\def\bd{\begin{Def}}
\def\ed{\end{Def}}
\def\bt{\begin{theorem}}
\def\et{\end{theorem}}
\def\bl{\begin{lemma}}
\def\el{\end{lemma}}
\def\bp{\begin{proof}}
\def\ep{\end{proof}}

\newcommand{\be}{\begin{equation}}
\newcommand{\ee}{\end{equation}}
\newcommand{\ba}{\begin{eqnarray}}
\newcommand{\ea}{\end{eqnarray}}

\def \bes {\begin{eqnarray}}
\def \ens {\end{eqnarray}}

\newcommand{\beq}[1]{\begin{equation}\label{#1}}
\newcommand{\eeq}{\end{equation}}

\everymath{\displaystyle}

\newcommand{\id}{\mathds{1}}

\begin{document}
\begin{cbunit}

\begin{titlepage}
\title{\vspace{-3cm} 
\includegraphics[height=0.3\textwidth]{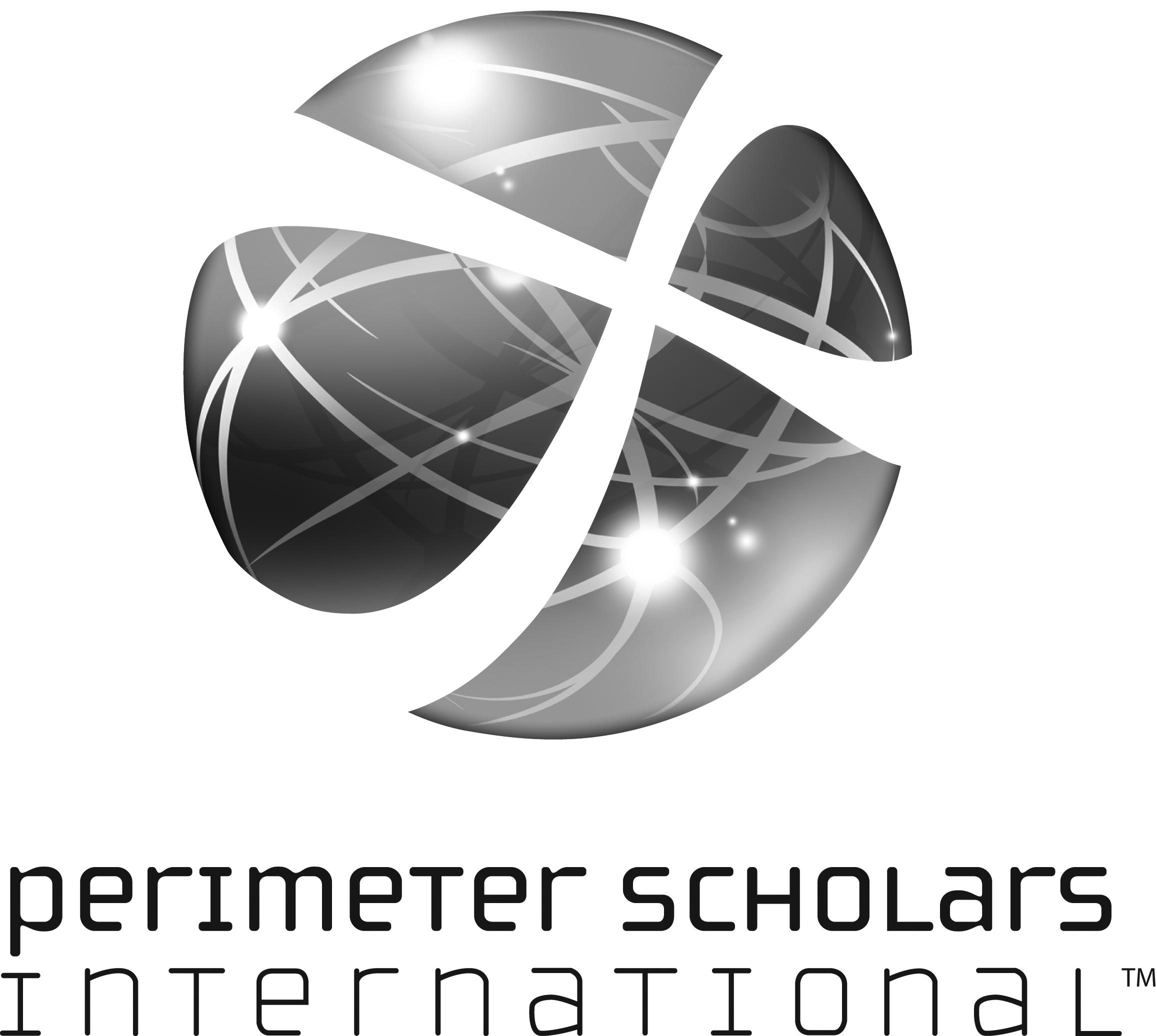}\\[2cm] 
 \textbf{A generalized multi-scale entanglement renormalization ansatz \newline \Large{More accurate conformal data from a critical lattice model}}\\[1cm]} 
 \author{\textbf{Javier Arg{\"u}ello Luengo}\\[2cm] 
 \textbf{
Supervised by Prof. Guifr{\'e} Vidal \& Dr. Ashley Milsted }\\[1cm]
 \textbf{An essay submitted}\\ 
 \textbf{for partial fulfilment of}\\ 
 \textbf{Perimeter Scholars International}\\[1cm]}
 \date{\textbf{June, 2017}} 
\maketitle 
\end{titlepage} 
 
\setcounter{page}{1}  

\chapter*{A generalized multi-scale entanglement renormalization ansatz\newline \huge{More accurate conformal data from a critical lattice model}} 
\chaptermark{Extracting scaling dimensions from generalized versions of MERA} 
\addcontentsline{toc}{chapter}{PSI Essay Template \emph{by Javier Arg{\"u}ello Luengo}} 
\setcounter{section}{0} 
\setcounter{equation}{0} 
\setcounter{table}{0} 
\setcounter{figure}{0} 

\vspace{-0.5cm} 
\centerline{ \textbf{Javier Arg{\"u}ello Luengo}} 
\vspace{0.3cm} 
\centerline{Supervisors: Prof. Guifr{\'e} Vidal \& Dr. Ashley Milsted} 
\centerline{\emph{Perimeter Institute for Theoretical Physics, Waterloo, Ontario N2L 2Y5, Canada}} 
\vspace{0.5cm} 


\begin{quote}
The multi-scale entanglement renormalization ansatz (MERA) provides a natural description of the ground state of a quantum critical Hamiltonian on the lattice. From an optimized MERA, one can extract the scaling dimensions of the underlying conformal field theory. However, extracting conformal data from the ascending superoperator derived from MERA seems to have a limited range of accuracy, even after increasing the bond dimension. Here, we propose an alternative ansatz based on an increasing number of disentangling layers. This leads to generalized versions of MERA that improve the extraction of scaling dimensions, as tested in the Gaussian MERA setting for the critical Ising model. 
\end{quote}


\section{Introduction}
Over the last decade, the multiscale entanglement renormalization ansatz has proved to be a useful tool for extracting conformal data from critical quantum systems~\cite{Pfeifer2008}. MERA originates from a better understanding of how previous techniques based on tensor networks work from the point of view of quantum information and unveils a connection with conformal field theories. 

The advantage of the renormalization group approach lies in the simultaneous exploration of many length scales. Widely used methods like Monte Carlo or statistical averaging find their limitations when dealing with increasingly small length scales, as the total number of grid points quickly reaches computer limits. In contrast, the renormalization group method uses one step for each length scale, integrating out fluctuations in sequence, from smaller to larger scales.


Following this spirit, MERA was originally conceived by Vidal to simulate critical phenomena~\cite{Vidal2007a}, subsequently proving to be a successful ansatz to describe ground states at criticality~\cite{Cincio2008,Evenbly2010a,Evenbly2010b}. In particular, it improves on some of the real-space renormalization ideas developed in the 1950s by Kadanoff, which were later developed by Kenneth Wilson in 1971 in the renormalization group theory~\cite{Wilson1983}. It is then natural to benchmark the connection between MERA and conformal field theories. In particular, scaling dimensions can be directly extracted from the diagonalization of one level of MERA optimized for the ground state of the theory, providing strong evidence of its behaviour as a renormalization scheme. Implications of these ideas have generated huge interest in other areas, including the AdS/CFT correspondence~\cite{Ryu2006} in holography~\cite{Bousso2002}, or the understanding of emergent topological order~\cite{Chang2013,Buerschaper2010} and frustrated antiferromagnetism~\cite{Harada2012,Lou2012} in condensed matter.

From a practical point of view, we are dealing with the contraction of a tensor network. One simple way to think about these structures is as a network of individual elements (the tensors), which are ``glued" together using entanglement~\cite{Orus2013}. Throughout the figures of this essay, tensor networks are shown using a diagrammatic representation due to Roger Penrose~\cite{Penrose1971}. There, each tensor is represented as a polygon and each index as a leg, so that bonds between different tensors describe the contraction of indices (see for example Figure~\ref{fig:disentangling}). The number of values each index can take is referred as the \textit{bond dimension}.

The finiteness of the bond dimension effectively limits the number of quasi-primary fields of the theory that can be observed. In practice, it turns out that fields with small scaling dimensions are mainly retained~\cite{Pfeifer2008}. In this work, we benchmark alternative ans{\"a}tze where isometries present a bigger causal cone (i.e., they see a larger region of the system), obtaining a more accurate description of higher scaling dimensions. 

The structure of the essay is as follows. In Section~\ref{sec:MERA}, we start reviewing the main ideas behind MERA, focusing on the importance of removing short-range entanglement and the application of disentanglers. These ideas lead to the generalized versions of MERA introduced in Sec.~\ref{sec:generalized}. In order to test the new ansatz, we apply it to the extraction of conformal data from a critical Ising model. The main concepts about conformal field theories and a description of how to extract scaling dimensions from MERA are introduced in Sec.~\ref{app:scaling}. 

In order to become a successful ansatz for criticality, we should compare its performance to the traditional scheme described by MERA. For this purpose, we should specify two main quantities. First, the computational requirements that each approach requires to evaluate the expected value of local observables (see Sec.~\ref{sec:compCost}). The evaluation of local Hamiltonians is needed during the optimization algorithm described in Sec.~\ref{sec:optimization} that allows one to get closer to the ground state of the critical theory. In particular, we shall focus on how the computational cost scales compared to the traditional strategy of increasing the bond dimension. Second, a figure of merit that allows us to compare the proficiency of both ans{\"a}tze. In Section~\ref{sec:results}, we analyze the difference in energies to the ground state and the accurate extraction of scaling dimensions smaller than or equal to three, discussing the difference in performance. 

\section{The Multiscale Entanglement Renormalization Ansatz}
\label{sec:MERA}
The goal of this section is to introduce the main concepts and ideas about the Multiscale Entanglement Renormalization Ansatz (MERA) that will be generalized in the following sections. A pedagogical and more complete overview on MERA and its applications can be found in~\cite{Hauru2013a}. 

\begin{figure}[b!]
\begin{center}
\includegraphics[width=\columnwidth]{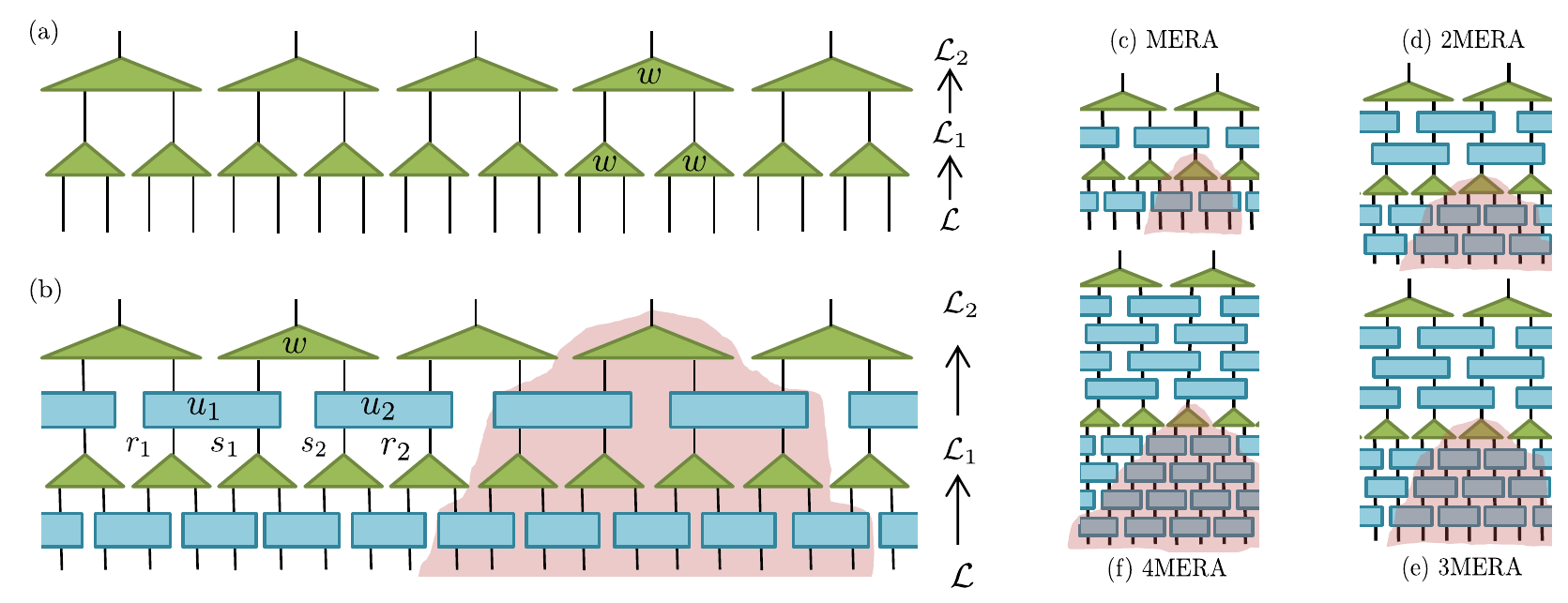}is\caption{\label{fig:disentangling} (a) Two lowest layers of isometries $w$ in a coarse-graining scheme where two sites are mapped into one. In particular, the initial lattice $\mathcal{L}_0$ is mapped into increasingly coarse-grained lattices $\mathcal{L}_1$ and $\mathcal{L}_2$. (b) Building upon the previous scheme, an additional layer of disentanglers allows us to remove short-range entanglement. This is the ansatz known as MERA. This idea is further generalized in the new versions of MERA represented in (d-f) that we explore throughout this essay. Shaded areas indicate the causal cone of some of the isometries.}
 \end{center}
\end{figure}
As we mentioned in the introduction, MERA belongs to a family of powerful techniques aimed at exploring the low-energy sector of critical Hamiltonians. It is based on a renormalization scheme, in which a lattice $\mathcal{L}$ is mapped to a coarse-grained lattice $\mathcal{L}'$, corresponding to a larger region of the system and effectively ``zooming-out" the lattice. This mapping to bigger scales is performed by \textit{isometries} $(w)$, which satisfy the relation,
\begin{equation}
\label{isom}
w^\dagger w = \id \,,
\end{equation}
so that,

\begin{equation}
\mathcal{L}_0 \xrightarrow{w_0^\dagger} \mathcal{L}_1 \xrightarrow{w_1^\dagger} \cdots \xrightarrow{w_{T-1}^\dagger} \mathcal{L}_T \,,
\end{equation}
where $\mathcal{L}_{n+1}$ is the coarse-graining of lattice $\mathcal{L}_n$ and the top lattice $(\mathcal{L}_T)$ is small enough to admit exact numerical computations. 

We should observe, however, that coarse-graining implies a loss of information. To give an intuitive example, we can think of the coarse graining of three spins into one, which we choose to be the alignment that satisfy the majority rules. As we can see, we are mapping eight possible states into two (up and down) and there is no possibility of knowing which was the exact initial situation if we only know that the coarse-grained spin is up. This property is reflected in the non-injectivity of $w^\dagger$ and non-surjectivity of $w$. Then, the operation of coarse-graining and then fine-graining the state does not necessarily return the initial input $(w w^\dagger \neq \id)$. 

The usefulness of this approach will depend on how computationally efficient it is to store the new state. As we will soon clarify with an example, this depends on how well we can characterize the ground state properties by keeping the largest eigenvalues of the new reduced density matrix, and disregarding the contributions from its eigenvectors with smaller eigenvalues. The dimension of this new state that needs to be stored is known as bond dimension $(\chi)$, and the smaller it is, the more useful the method will become.

However, $\chi$ quickly grows in important situations, like criticality, preventing it from being computationally useful. Intuitively, short-range entanglement is not integrated out by coarse graining and its accumulation makes the dimension of the Hilbert space increase. It also carries non-physical artifacts that depend on the specific renormalization scheme, and is therefore undesirable.

One simple example\footnote{This example is originally ilustrated in~\cite{Cincio}.} we can think about is having four spin-1/2 particles in a chain and coarse-graining the two intermediate spins $(s_1,\, s_2)$, (see Figure~\ref{fig:disentangling}). In particular, we take the state,
\begin{equation}
\ket{\psi}=\co{\frac{1}{\sqrt{2}}\pa{\ket{01}+\ket{10}}}^{\otimes 2}\,.
\label{state}
\end{equation}
Looking at its reduced density matrix on $s_1,\,s_2$, we see that it is a maximally mixed state,
\begin{equation}
\rho_{s_1s_2}=\text{Tr}_{r_1r_2}\ket{\psi}\bra{\psi}=\frac{1}{4}\pa{\ket{00}\bra{00}+\ket{11}\bra{11}+\ket{01}\bra{01}+\ket{10}\bra{10}}\,.
\end{equation}
In this situation, four parameters are needed to describe the state, and it is not possible to truncate the effective coarse-grained state without losing a considerable amount of the information.

To overcome these issues, Vidal proposed the concept of \textit{entanglement renormalization}~\cite{Vidal2007a, Evenbly2010}. The main idea is astonishingly simple. As the accumulation of entanglement originates a limitation in the truncation of the effective final space, it is worth removing short-range entanglement from the system before applying the coarse-graining, so that it does not accumulate, as used to happen in previous methods. This allows us to both consider arbitrarily large lattice systems and to restore the scale invariance that is characteristic of critical phenomena~\cite{Giovannetti2008, Pfeifer2008}. This is achieved by including layers of unitary transformations that are able to remove entanglement at short scales, being known as \textit{disentanglers} $(u)$. Unlike isometries, they preserve the dimension of the state space where they operate, and define a bijective operation,
\begin{equation}
\label{disen}
uu^\dagger = u^\dagger u = \id\,.
\end{equation}

Going back to the example of a maximally mixed state, we can now see how useful disentanglers are. In particular, we choose an operation that transforms the state $\frac{1}{\sqrt{2}}\pa{\ket{01}+\ket{10}}$ into the state$\ket{00}$. This can be done with the unitary transformation,
\begin{equation}
u=\sum_{abij}u_{ij}^{ab}\ket{a}_r\ket{b}_s\bra{i}_r\bra{j}_s\,,
\end{equation}
with $u^{00}_{01}=u^{00}_{10}=u^{01}_{01}=-u^{01}_{10}=\frac{1}{\sqrt{2}}$; $u^{10}_{00}=u^{11}_{11}=1$ and the remaining elements equal to zero. Applying this transformation to the state in Eq.~\eqref{state}, we  obtain a new reduced density matrix which is simply given by,
\begin{equation}
\tilde{\rho}_{s_1s_2}=\ket{00}\bra{00}\,.
\end{equation}

We can see that three of the eigenvalues are now null, and only a non-zero element is needed to describe the new reduced state. Then, an appropriate choice of disentangler has allowed to reduce the dimension of the effective block from its maximum value, four, to the minimum possible dimension, one.

\subsection{Generalized versions of MERA}
\label{sec:generalized}
Based on the idea of adding disentanglement, several approaches can be used to improve the results of the ansatz, and each of them might be more adequate for specific purposes. One common strategy consists in simply increasing the bond dimension of the ansatz. By dealing with more degrees of freedom, a bigger region of the system is seen by each tensor and entanglement between further regions can be removed by individual disentanglers.

Alternatively, in this essay we propose to use generalized versions of MERA where additional layers of disentanglers are added. In particular, we denote as $n$MERA the generalized version of MERA where each level\footnote{In order to avoid confusion, \textit{level} refers to the combination of disentaglers+isometry that form a coarse-graining operation. We reserve the term \textit{layer} to denote each line of disentanglers or isometries in the diagram.} is composed by $n$ layers of disentanglers and one layer of isometries.

When compared to increasing the bond dimension of the ansatz, these generalized versions of MERA present some advantages. First, they offer a more compact description of the ascending superoperators, as less variational parameters are involved\footnote{In Table~\ref{tab:scaling} we can see that the space cost grows quadratically in the bond dimension, but only linearly in the number of layers of disentanglers.}. Second, the causal cone of each isometry grows, as we can appreciate in Fig.~\ref{fig:disentangling}. This allows us to remove short-range entanglement at higher distances before applying coarse-graining. Third, it enables to build a sequential algorithm to improve the solution. For example, we can build upon previous results obtained for less layers of disentanglers and reoptimize after the addition of one layer of identities. We will explain this strategy in more detail in section~\ref{sec:results}.

To benchmark this new ansatz, we focus on the extraction of conformal data. Previous results on MERA~\cite{Pfeifer2008} successfully retain conformal fields with small scaling dimensions. However, the authors report that it is challenging to obtain this information for higher scaling dimensions, even after an increase of the bond dimension. Alternatively, we want to explore if the addition of more layers of disentanglers in the ansatz improves the extraction of scaling dimensions. 

\section{Scaling dimensions and the critical Ising model}
\label{app:scaling}
\begin{figure}[t!]
\begin{center}
\includegraphics[width=\columnwidth]{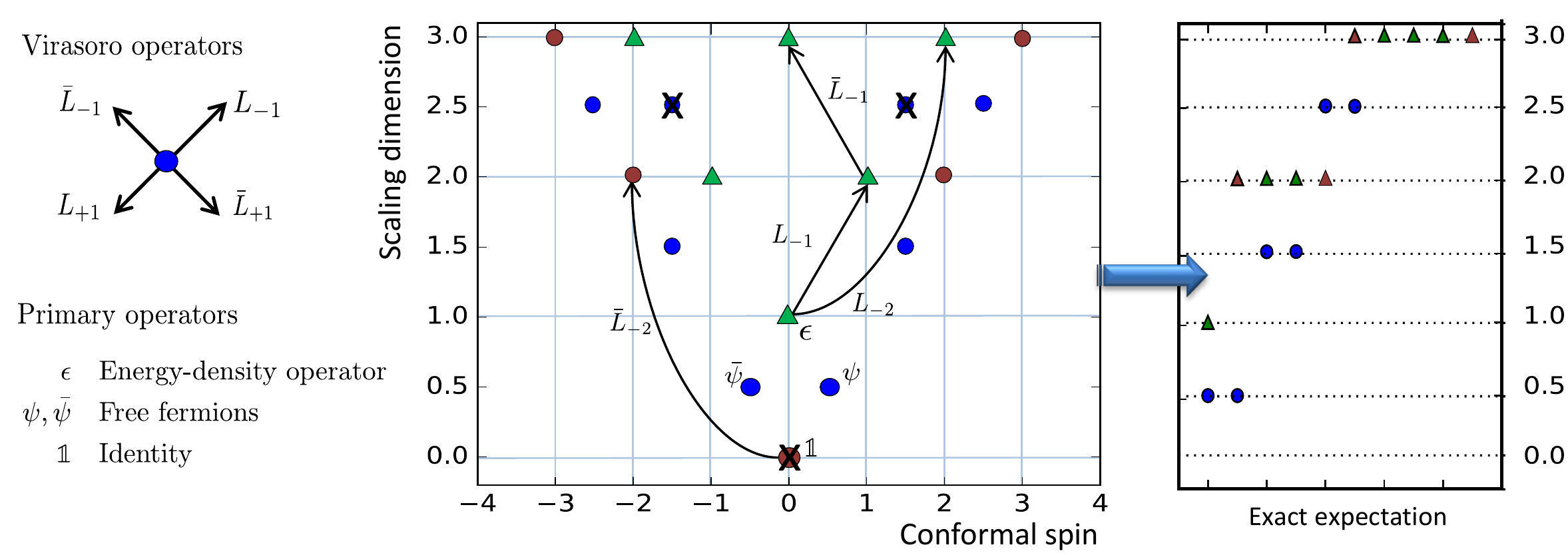}
\caption{\label{fig:virasoro} \textit{Left.} Scaling dimensions and conformal spin corresponding to scaling operators present in the critical Ising model. Primary operators [identity ($\id$), free fermions ($\psi$, $\bar{\psi}$) and the energy-density operator ($\epsilon$)] are independently labeled and its descendants are indicated with the same marker. The action of some of the ladder operators are represented as an example. \textit{Right.} Scaling operators that are linear or quadratic in fermionic operators are diagrammatically represented as shown in Sec.~\ref{sec:results}. Note on the diagram that the degeneracies we expect to observe are $0-2-1-2-4-2-5$.}
\end{center}
\end{figure}
In Section~\ref{sec:results}, we want to compare the numerically observed degeneracy of the scaling dimensions with the  expectation for the subjacent field theory. Here, we briefly introduce some of the theoretical tools that allow us to extract this data. A complete overview about this topic can be found in~\cite{Henkel}, Chapters 4 and 10.

The structure of a conformal field theory is characterized by its scaling operators, $\varphi_\alpha(z)$. They transform covariantly under scale transformations or rotations. This means,
\begin{align}
z \to \mu z & \Leftrightarrow  \varphi_\alpha(0) \to \mu^{-\Delta_\alpha} \varphi_\alpha(0)\,,\\
z \to \text{e}^{i\theta} z & \Leftrightarrow  \varphi_\alpha(0) \to \text{e}^{-i\theta S_\alpha} \varphi_\alpha(0)\,.
\label{eq:conformal}
\end{align}
where $\Delta_\alpha$ and $S_\alpha$ denote the \textit{scaling dimension} and \textit{conformal spin} of $\varphi_\alpha(z)$. The set of all scaling operators is completely defined by some basic building operators called \textit{primary}. Primary operators completely fix the structure of the theory and can generate the infinite set of descendant operators, organized in \textit{Virasoro towers}.

For the Ising quantum chain discussed in this essay, four primary operators appear: the identity ($\id$), the free fermions ($\psi$, $\bar{\psi}$), and the energy-density operator ($\epsilon$). Each of them has its own conformal tower, revealed by the application of the Virasoro generators. These generators define an algebra described by the action of the energy-momentum tensor on physical scaling operators, following, 

\begin{equation}
L_n=\frac{1}{2\pi i} \oint dz\, z^{n+1} T(z)\,.
\label{eq:Virasoro}
\end{equation}

The Virasoro towers associated to each of the four primary fields are represented in Fig.~\ref{fig:virasoro}. For instance, we observe that there is no descendants of the identity with scaling dimension $\Delta_i=1$ because of the regularity of integral~\eqref{eq:Virasoro} associated to $\L_{-1}\id$. This situation changes with the non-vanishing operator $\L_{-2}\id$, that gives the energy-momentum tensor $T$, using the Residue Theorem. 

Following the numerical approach discussed in section~\ref{sec:scaling}, we will focus on extracting scaling dimensions of operators which are linear or quadratic in fermionic operators. It is then important to note the change in the number of fermionic operators present in the components of these towers. While $\id$ is order zero in fermionic operators, its descendant, $\L_{-2}\id=T \propto \psi \partial \psi$~\cite{DiFrancesco1997} is quadratic and should appear among the extracted scaling dimensions. Following the same argument, we do not expect to see the descendants $\L_{-2}\bar\psi,\; \bar\L_{-2}\psi $, as they now involve three fermionic operators. These arguments guide the exact expectation schematized in Figure~\ref{fig:virasoro} that we wish to compare to the numerical results obtained using MERA.

\subsection{Extracting scaling dimensions from MERA}
\label{sec:scaling}
A coarse-graining transformation for MERA has recently been pushed forward~\cite{Evenbly2014a}. In this strategy, each of the layers of the ansatz can be seen as a coarse-graining step that allows us to gradually extract short-range entanglement. This naturally leads to scale-invariance which is made explicit when exploring systems at criticality, i.e. when its properties are not affected by zooming-in or zooming-out the system. 

As mentioned before, we want to explore if the addition of more layers of disentanglers leads to a more accurate description of the scaling dimensions of descendant operators at criticality. In particular, we shall focus on the free theory described by the critical Ising Hamiltonian. Within the Gaussian formalism for free theories~\cite{Bravyi2004}, we can easily coarse-grain states which are linear or quadratic in fermionic operators (see Figure~\ref{fig:ascending} for a description of this method). In this case, we can see the ascending superoperator as a linear map that transforms a state into the corresponding coarse-grained state. In order to be well defined, this transformation must map a certain number of lattice sites $k$ into the same number of sites. Otherwise, the transformation could not be applied again to the new state. As we can see on the right of Figure 4, this value corresponds to $k=3$ in MERA. When we generalize this approach to $n$MERA we observe that the minimum value we can take obeys the rule $k=2n+1$, increasing with $n$ due to the growth of the causal cone. The ascending superoperators corresponding to 2MERA $(k=5)$ and 3MERA $(k=7)$ are shown in the right of Figure~\ref{fig:computational}.

\begin{figure}[t!]
\begin{center}
\includegraphics[width=0.8\columnwidth]{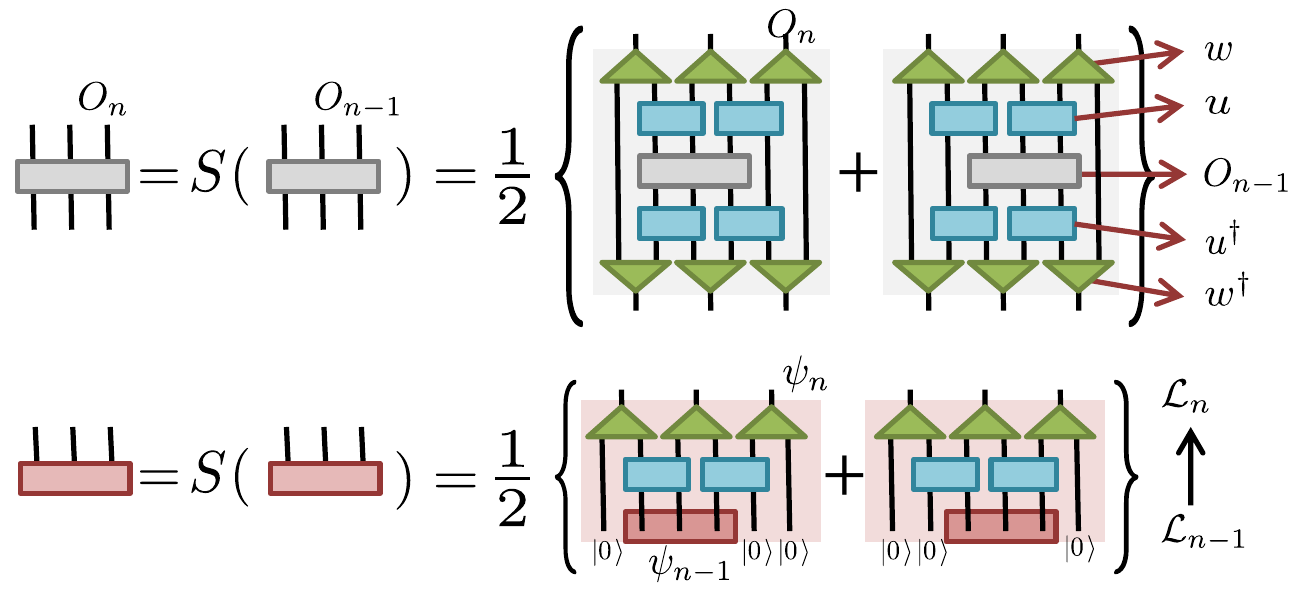}
\caption{\label{fig:ascending} \textit{Bottom.} Schematic representation of the ascending superoperator $S$ that maps a state $\psi_{n-1}$ (shown in red) in the lattice $\mathcal{L}_{n-1}$ to the ascended state $\psi_n$ in the coarse-grained lattice $\mathcal{L}_{n}$. \textit{Top. }Analogously, we show the ascension of the 3-site operator $O_{n-1}$ shown in grey, which leads to an operator supported on the same number of sites.}
\end{center}
\end{figure}

The corresponding scaling dimensions of the theory can be extracted from the eigenvalues of this liner map. To have some intuition of how this is the case, we observe that a binary MERA transforms $N$ sites into $N/2$ sites. From the point of view of a CFT, a dilation, $x\to z'= 2^{-1} z\,,$ corresponds to a conformal transformation (see Eq.~\eqref{eq:conformal}),
\begin{equation}
\varphi_\alpha \to \varphi'_\alpha= 2^{\Delta_\alpha}\varphi_\alpha=\lambda_\alpha \varphi_\alpha\,.
\end{equation}
With this approach, we are able to extract the scaling dimension associated to operators that are linear or quadratic in fermionic operators by simply calculating
\begin{equation}
\Delta_\alpha=\log_2 |\lambda_\alpha|\,.
\end{equation}

This result directly shows the connection between the coarse-graining applied at each level of MERA and the associated conformal field theory.

\section{Optimization algorithm}
\label{sec:optimization}
In order to follow Sec.~\ref{sec:scaling} to extract scaling dimensions, we firstly need to optimize the ansatz for the ground state of the critical quantum Ising model. For this purpose, we choose the energy of the state as the cost function to be minimized and follow the strategy proposed in~\cite{Evenbly2009}.  The main idea behind is that each tensor of the ansatz is individually optimized to minimize the global energy of the state. It also takes advantage of the translational and scale invariance of the Hamiltonian, reducing the computational cost required in the optimization. In order to keep the flow of this essay, we review it in more detail in Appendix~\ref{app:optimization}. 

One of the difficulties we need to face when dealing with generalized versions of MERA is that  more variational parameters are found in the system as the number of layers of disentanglers increases. Then, it appears to be more difficult to obtain good approximations to the ground state starting from randomly initialized tensor elements. To face this issue, we propose a systematic method to increase the number of layers of disentanglers starting from optimized solutions for the system with one less layer of disentanglers. This is, to initialize $n$MERA from a previous solution for $(n-1)$MERA.  The main idea, schematized in Figure~\ref{fig:growing}, is based on the addition of layers of identities at the bottom of each level. This preserves the previous solution, while adding more degrees of freedom when considering the expanded version of MERA as the initial state to be optimized.

\subsection{Computational cost}
\label{sec:compCost}
Even if these are more suitable ansatz to explore criticality, we are limited in practice by the computational cost required to compute these structures. This is, the number of operations the computer needs to perform in order to contract the basic blocks, These building blocks are needed to calculate superoperators and the expected value of the Hamiltonian that appears in the cost function. In order to become a useful method, the contraction of generalized versions of MERA should be comparable in cost to the usual approach based on increasing the bond dimension of the ansatz. 

\begin{figure}[t!]
\begin{center}
\includegraphics[width=\columnwidth]{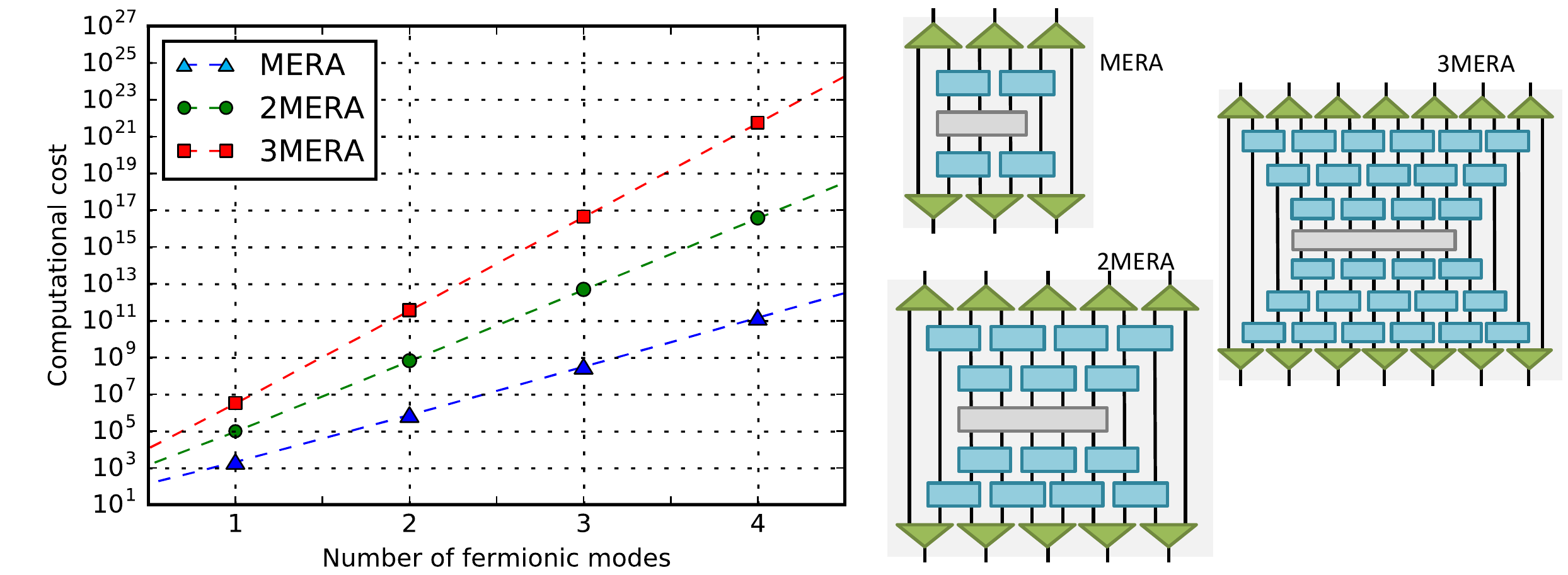}
\caption{\label{fig:computational} Computational cost of the contraction of the ascending superoperators corresponding to MERA, 2MERA and 3MERA as a function of the number of fermionic modes $\chi=2^{N_f}$. On the right, we represent the application of the ascending superoperators corresponding to MERA, 2MERA and 3MERA on the local $(2n+1)$-site observables shown in grey.}
\end{center}
\end{figure}

When contracting the diagram, the computational cost will highly depend on the order in which contractions are performed, and identifying the optimal order is not an obvious task. Using the Netcon algorithm proposed by Pfeifer, Haegeman and Verstraete~\cite{Pfeifer2013}, we can obtain the computational cost for the ascending operators of the different ans{\"a}tze, which we summarize in Table~\ref{tab:scaling}. 

Importantly, we should note that the critical Ising model is a free theory and can be diagonalized as indicated in Appendix~\ref{app:ising}. One of the main consequences of this is that $N$ fermionic modes suffice to characterize the $2^N$-dimensional Hilbert space required for interacting theories. From now on, we define $N_f$ as the number of fermionic modes encoded per leg.

If we think of isometries as a unitary transformation where one input is fixed, we can now see MERA as a quantum circuit for the inputs. From this perspective, we are interested in discussing the situation where few fermionic modes are considered per leg. This scenario is closer to the intuitive idea of a quantum circuit in which each isometry encodes an individual mode of the system. For example, in the limiting case of encoding one fermionic mode per leg, excited states of the system can be found by independently switching the occupation of the different fermionic modes, recovering the full spectrum of the model used to optimize MERA.

In Figure~\ref{fig:computational}, we plot the computational cost of contracting the generalized versions of MERA for low values of fermionic modes. As an example, we observe that contracting a 3MERA with 2 fermionic modes per leg (corresponding to $\chi=4$ in interacting theories) is computationally similar $(\approx 10^{11}$ operations$)$ to the contraction of a MERA with a bond dimension of $\chi=16$ $(N_f=4)$. The question we now want to address is which of both approaches provide a more accurate description of the conformal data while paying a similar computational cost\footnote{Note that we are looking for a comparison which is also fair for interacting theories, even though this essay is focused on free theories.}.

\begin{table}
\begin{center}
\begin{tabular}{ cccc } 
\hline
&Computational cost& \multicolumn{2}{c}{Computational cost (Free)} \\
Ansatz & (Interacting) & Time cost  & Space cost  \\
\hline
MERA & $2\chi^9 + 4\chi^8  + 2\chi^6 + 2\chi^5 $ & $\ord{N_f^3}$ & $\ord{N_f^2}$\\ 
2MERA & $8\chi^{13} + 8\chi^{12} + 6\chi^6 + 2\chi^5$ & $2\ord{N_f^3}$ & $2\ord{N_f^2}$\\ 
3MERA & $18\chi^{17} + 15\chi^{16} + 10\chi^6 + \chi^5$  & $3\ord{N_f^3}$ & $3\ord{N_f^2}$\\ 
\hline
\end{tabular}
\caption{\label{tab:scaling} Computational cost for the contraction of the ascending operator for the explored structures. In the interacting theory, the bond dimension $(\chi)$ corresponds to $2^{N_f}$, where $N_f$ stands for the number of fermionic modes encoded per leg. In the Gaussian formalism, contractions reduce to a product of matrices, which is subleading when compared to the diagonalization of the tensor elements. Time cost scales as the cube of the size of these matrices, while the space cost scales as the number of components.} 
\end{center}
\end{table}

\section{Results}
\label{sec:results}
\begin{figure}[t!]
\begin{center}
\includegraphics[width=\columnwidth]{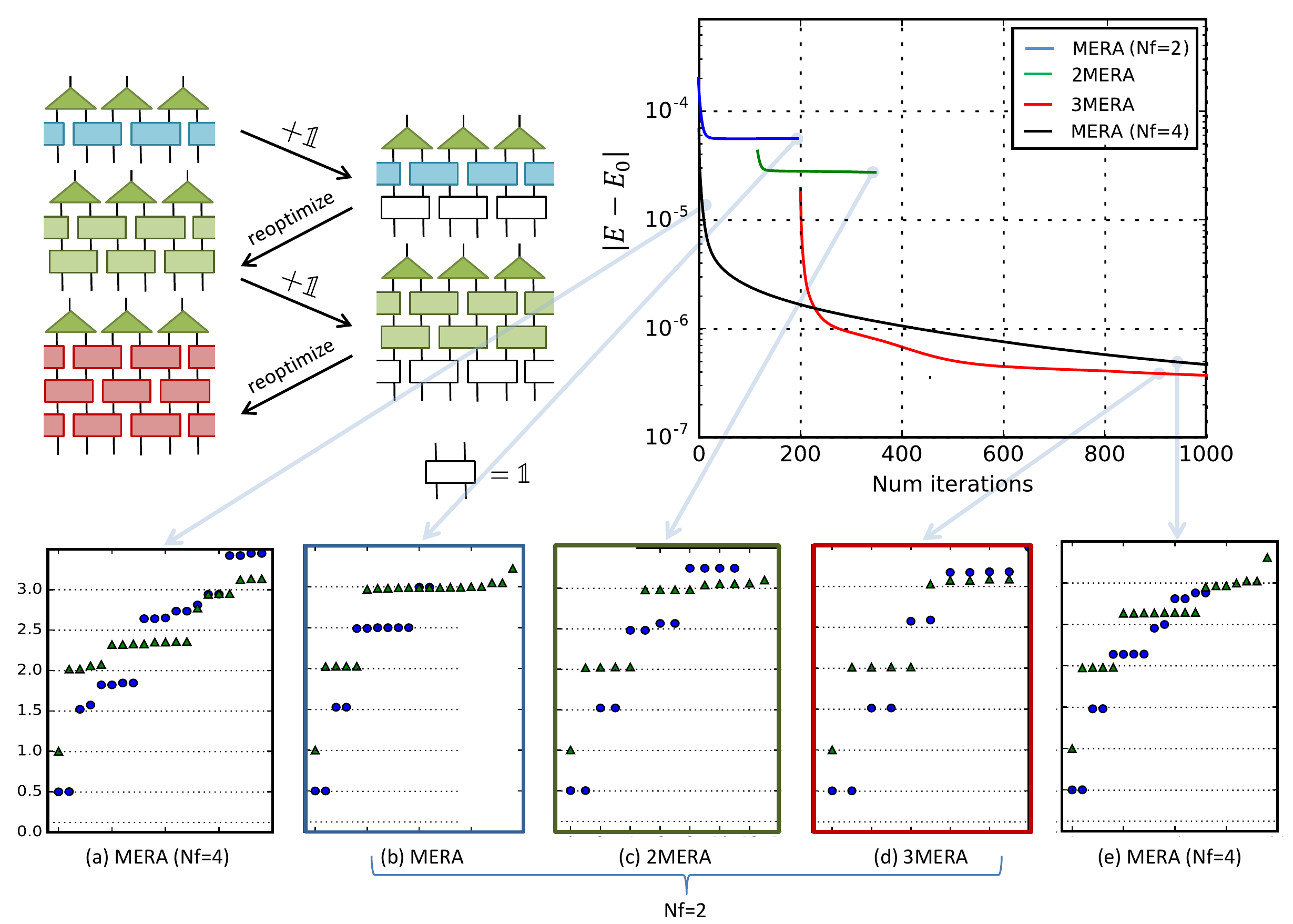}
\caption{\label{fig:growing} \textit{Top. }Difference in energies for the analytic ground state of the Ising critical model following the strategy schematized on the left: a layer of identities is added to an initially optimized MERA. The resulting 2MERA is effectively equal to the previous solution, but can now be reoptimized making use of the additional degrees of freedom that these disentanglers provide. The process is repeated once more to get an optimized 3MERA and the resulting structure is compared to the performance of MERA with $N_f=4$. \textit{Bottom. }Scaling dimensions associated to linear and quadratic operators are extracted at key points of the optimization process. The horizontal axis just allows us to appreciate the degeneracy of each level.}
\end{center}
\end{figure}
	
In this section, we benchmark the effect of our proposed ansatz, $n$MERA, in the extraction of scaling dimensions for the critical Ising model. Following the optimization approach described in Sec.~\ref{sec:optimization}, we start optimizing a randomly initialized MERA with two fermionic modes per leg. The strategy described in Fig.~\ref{fig:growing} allows one to systematically grow this solution to reach the optimized 3MERA that we use to extract scaling dimension.

\begin{figure}[t!]
\begin{center}
\includegraphics[width=\columnwidth]{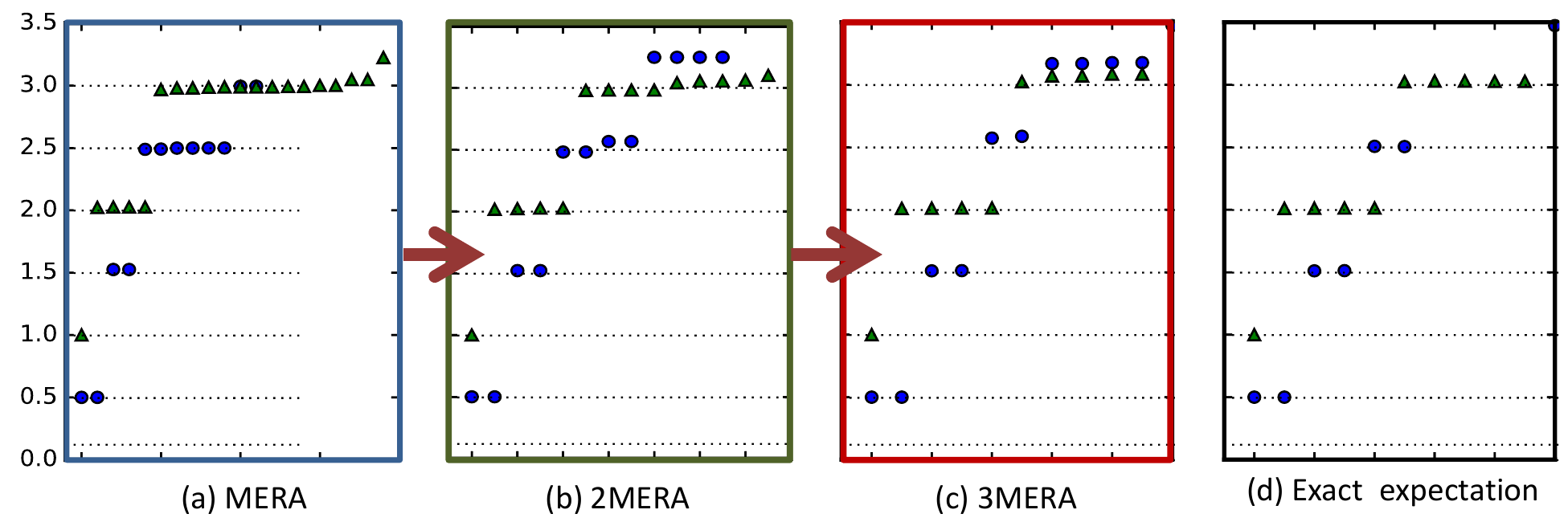}
\caption{\label{fig:result} (a-c) Scaling dimensions for the critical Ising model extracted from generalized ans{\"a}tze with an increasing number of layers of disentanglers following the setup and strategy described in Figure~\ref{fig:growing}. (d) Exact scaling dimensions expected for the model following Section~\ref{app:scaling}.}
\end{center}
\end{figure}

In order to make a fair comparison with MERA, we want to relate ans{\"a}tze which are similar from the computational point of view. In Section~\ref{fig:computational}, we showed that the computational cost of contracting a 3MERA with two femionic modes per leg is comparable to a regular MERA with $N_f=4$, i.e., a bond dimension four times bigger. If getting an approximation of the ground state energy is our main interest, we observe that both ans{\"a}tze approximate this value to the order of $10^{-6}$ after $1000$ iterations\footnote{This takes around $1$h using a 2 GHz laptop with 8 Gb of RAM.}. 

Remarkably, the performance is drastically different when we look at the scaling dimensions extracted at the different steps of the optimization. To compare the conformal data extracted from the generalized versions of MERA, we focus on the expected degeneracies of scaling dimension smaller or equal to 3 that, for the explored Ising critical model, correspond to 
$$0-2-1-2-4-2-5$$
(see Section~\ref{app:scaling}). When looking at traditional MERA, we consistently\footnote{In practice, 50 random initializations were considered, and Figure~\ref{fig:growing} represents the optimization of the initialization that reached the best approximation in energy to the ground state} observe that its standard version satisfactorily reproduces scaling dimension smaller than two, but overestimates the degeneracies for values bigger than the stress-energy operator. This scenario drastically changes when using the generalized versions of MERA. As the ansatz grows, we observe in Figure~\ref{fig:result} that the overcounting of operators observed in MERA tends to disappear,  reaching the analytic degeneracies for scaling dimensions smaller or equal to three, in the case of the 3MERA with only two fermionic modes per leg.

Despite the advantages of this strategy, one of the aspects we observe is that it leads to solutions where the disentanglers in the new layer remain very close to the identity. Although this is enough to improve the convergence in energies by an order of magnitude and reach the right degeneracies for the scaling dimensions, it might not be the best approach. Exploring alternative strategies that more efficiently share the disentangling power among the different layers might lead to even better results, and its investigation is left for future work.

\section{Conclusions}
In this essay we have explored how to extract conformal data from the ground states of quantum critical systems using new generalized versions of MERA where additional layers of disentanglers are included. 
Focused on the extraction of conformal data from the critical Ising model, we implement a new strategy that allows us to initialize the generalized versions of MERA building upon previously optimized solutions with fewer layers of disentanglers. Following this approach, generalized ans{\"a}tze provide better approximations to the ground state in terms of energies and scaling dimensions when compared to MERA with a similar computational cost. 

Further extensions of this work shall explore interacting systems, where the computational requirements of these ans{\"a}tze are a more important restriction, as the Gaussian formalism cannot be applied. Additional strategies to initialize the ansatz from previous solution would also be highly desirable. We finally note that most of the ideas behind the generalized versions of MERA are independent of the spacetime dimension and could also be explored in principle for critical ground states of 2D systems. 

\section{Acknowledgements}
I would like to firstly express my deep gratitude to my supervisors: Guifr{\'e} Vidal, for being an approachable, understanding and outstanding mentor; and Ashley Milsted for his continuous guidance during this project and generously sharing his time whenever I needed to discuss or the code didn't work. I am also indebtedly grateful to everyone involved in the PSI program, from Debbie, Erica and the fellows to all the PSI students with whom I have shared so many experiences and good moments. This year would have neither been the same without the Spanish team and this essay has highly benefited from their comments and careful reading of the manuscript. Finally, I would like to thank my family for keeping on taking care of me from the other side of the ocean, even when they didn't fully understand what all of this was about.


\newpage
\begingroup
\let\cleardoublepage\clearpage
\bibliographystyle{bibstyleNCM}

\endgroup

\clearpage

\begin{subappendices}
\renewcommand\thesection{\Alph{section}}

\section{Details of the optimization algorithm}
\label{app:optimization}

\begin{figure}[b!]
\begin{center}
\includegraphics[width=\columnwidth]{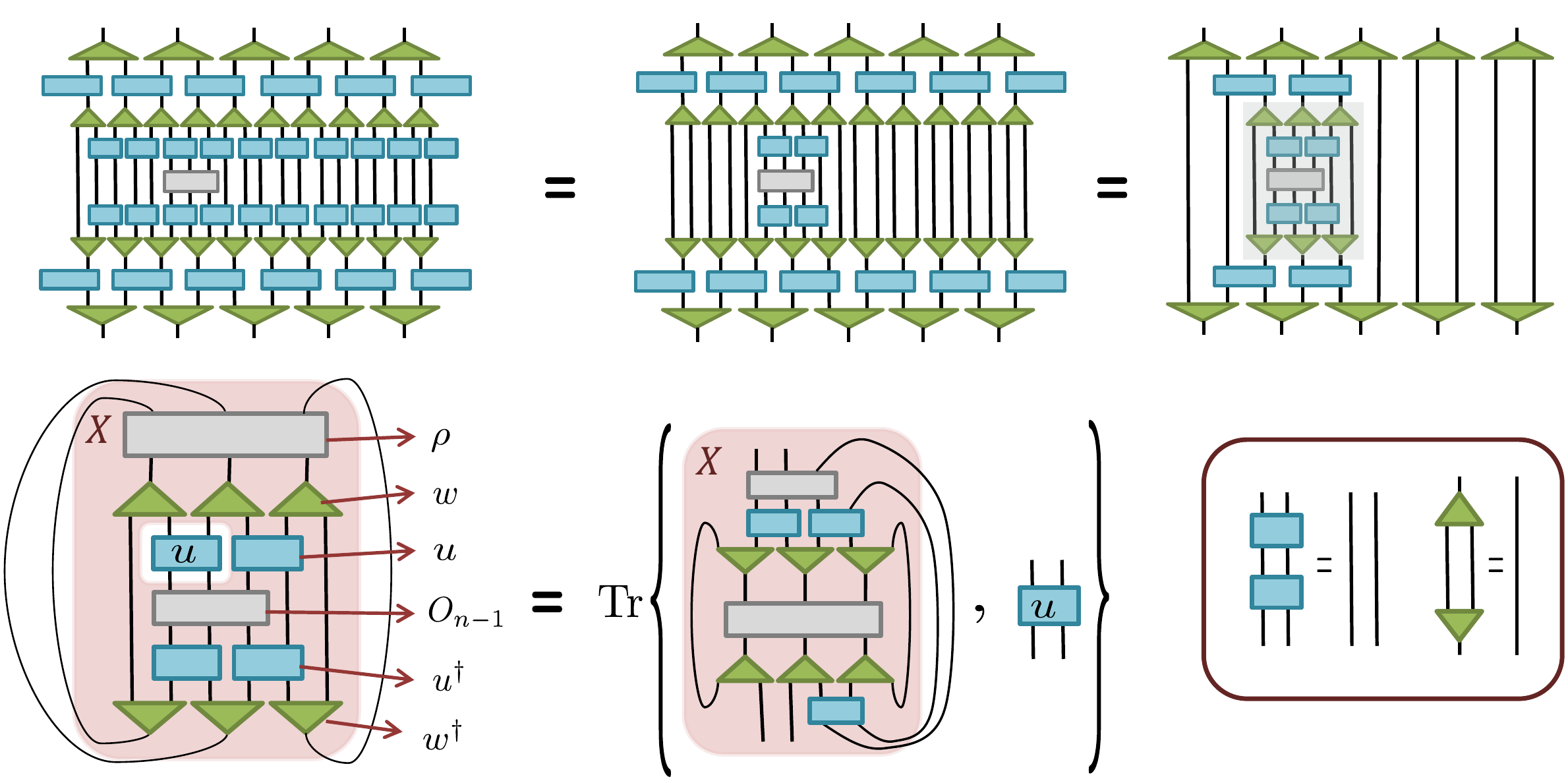}
\caption{\label{fig:optimization} The unitarity condition imposed to isometries (Eq.~\ref{isom}) and disentanglers (Eq.~\ref{disen}) can be diagrammatically represented as shown in the box. \textit{Top.} This representation allows us to easily visualize that the contraction of local terms with MERA can be drastically simplified, reducing an infinite MERA to the simple evaluation of the building blocks discussed in section~\ref{sec:compCost}. \textit{Bottom.} When optimizing a given disentangler, we split the contraction of the building block in two different parts: the disentangler ($u$), and its environment ($X$), shown in red. The singular value decomposition of this environment will lead to the updated version of the disentangler.}
\end{center}
\end{figure}
Here we review some of the details of the optimization. As we schematize in Figure~\ref{fig:optimization}, the evaluation of local observables is reduced to the contraction of specific finite building blocks that depends on the used $n$MERA. Each of the tensors in these building blocks needs to be individually optimized. To minimize its contribution in energies, we distinguish two regions: the tensor to be optimized $(A)$, and its environment $(X)$, which corresponds to the complementary part of the tensor network (see Figure~\ref{fig:optimization}). The total energy of the local Hamiltonian then reads as $\text{Tr}(AX)$. We want to update $A$ with a new tensor $A'$ that decreases the total trace. To minimize this quantity, we singular value decompose the environment as $X=USV^\dagger$, where $S$ is a diagonal matrix that contains the singular values of the tensor, and update the tensor as $A'=-VU^\dagger$. 

\begin{equation}
\min_A \text{Tr}(AX)= \min_A \text{Tr}(AUSV^\dagger)= -\text{Tr}(A'USV^\dagger)= - \text{Tr}(S)=-\sum_\alpha s_\alpha
\end{equation}
(here $s_\alpha \geq 0$ are the singular values of $X$).
It is important to note that the complexity of the singular value decomposition is below $\ord{\chi^8}$, and is then subleading.

Now that we can optimize individual tensors, we can explore the structure in an organized way. Different strategies can be followed, and we generalize the approach proposed by Evenbly and Vidal in~\cite{Evenbly2009} and follow a bottom-up approach. As a starting point, we initialize MERA with random tensors and optimize the structure from the bottom levels to the top. Following this approach, we firstly need to descend the top state to the lowest level. We do this sequentially. Then, the top state is descended  through the highest level to obtain the density matrix $\mathcal{L}_{N-1}$. This reduced density matrix can now be descended through the following level to get $\mathcal{L}_{N-2}$, and so on, until we reach the first level. As we can recycle the previous calculations, only a global contraction is needed per level.

The algorithm we use is the following. First, we start optimizing the first level of our $n$MERA. For each level, we optimize the first layer of disentanglers, then we move to the second one, and continue until we reach the corresponding layer of isometries. This in-level swiping is repeated $q_{level}$ times before moving to the next level, so that enough convergence is reached. One of the main advantages of this method is that small values of $q_{level}\approx 10$ are satisfactory in practice, and it is preferable not to insist on optimizing a level that may drastically change in the following swipe. Once we reach the top state, we optimize it by just using the diagonalization of the ascended state.

Now that we can optimize the ansatz, we can make the optimization more efficient by using two important properties of our quantum critical system: translational and scale invariance.
Due to the translational invariance of the Hamiltonian, all disentanglers in a given layer can be considered equal. Then, we jointly optimize them by averaging the possible environments that can appear, and consider it as the effective environment. 

Furthermore, the fact that the system is invariant under a change of scale draws a beautiful connection between criticality and the coarse-graining effect of the isometries. When seeing each level of MERA as a superascending operator, critical states will be mapped into themselves. Effectively, only one layer needs to be self-consistently optimized until reaching criticality, saving resources from the computational space perspective. From the practical point of view, we observe that including two transition layers of MERA between the Hamiltonian and this critical layer facilitates convergence.

\section{Diagonalization of the critical Ising model}
\label{app:ising}
The critical Ising model can be seen as a particular case of the more general XY model,
$$
H=-\sum_{n=1}^N\co{t\pau{z}{n}+\frac{1+\eta}{2}\pau{x}{n} \pau{x}{n+1}+\frac{1-\eta}{2}\pau{y}{n}\pau{y}{n+1}}\,,
$$
with periodic boundary conditions,
$\pau{x,y}{N+1}=\pau{x,y}{1}\,.$ For $\eta=1$, it reduces to the Ising model, which is critical for $t=1$. One of the advantages of this model is that it can be exactly diagonalized, allowing to compare our numerical results with exact expectation. In this appendix, we will briefly indicate the main steps involved in this diagonalization, so that we can appreciate the appearance of fermionic modes.

To begin with, it is useful to express the Hamiltonian as a quadratic expression in Majorana operators. That allows us to perform numerical calculations from the point of view of the Gaussian formalism for quadratic Hamiltonians, highly simplifying the computational performance~\cite{Bravyi2004}. This mapping between spin variables and free fermionic operators can be obtained using the Jordan-Wigner transformation,
\begin{equation}
c_l=\pa{\prod_{m<l}\sigma^z_m}\frac{\sigma_l^x-i\sigma_l^y}{2}\,, \quad c_l^\dagger=\pa{\prod_{m<l}\sigma^z_m}\frac{\sigma_l^x+i\sigma_l^y}{2}\,.
\end{equation}

Note that the Hamiltonian now becomes quadratic. It is now also useful to write the creation and annihilation operators in terms of the $2N$ Majorana operators\,,
\begin{equation}
\psi_{l,1}=c_l + c_l^\dagger \,, \quad \psi_{l,2}=i(c_l-c_l^\dagger)\,,
\end{equation}

which satisfy the properties,
$$
\psi_i^\dagger = \psi_i \,, \quad \lla{\psi_i, \psi_j}=2\delta_{i,j}
$$
In terms of Majoranas, the Hamiltonians can be expressed as a quadratic form,
$$
H=\frac{i}{2}\sum_{i,j}\psi_i C_{ij} \psi_j = \frac{i}{2}\vec{\psi}^ \dagger C \vec{\psi}\,,
$$
where $\vec{\psi}=\pa{\psi_{1,1} \quad
   \psi_{1,2} \quad
   \psi_{2,1} \quad
   \ldots	\quad
   \psi_{N,1} \quad
   \psi_{N,2}} ^T\,.$
   
Due to the hermiticity of the Hamiltonian, we can choose the matrix $C$ to be real $(C=C^*)$ and skew-symmetric $(C^T=-C)$. This allows us to decompose the quadratic form as, $C=O\tilde{C}O^T$, with $O$ an orthogonal matrix and $\tilde{C}$ a matrix in the form $\tilde{C}=\oplus_{k=1}^N \pa{\begin{array}{cc}
   0 & \epsilon_k \\
   -\epsilon_k & 0 \\
  \end{array}}\,,$ being ${\epsilon_k}$ real values.

While we are focusing on describing the main steps in the diagonalization, the particular expression for the Ising model can be checked in~\cite{Henkel}. The Hamiltonian can be finally expressed as a sum of number operators, which can be built from Majorana modes,
$\vec{\phi} = O^\dagger \vec{\psi}$, as $b_k=\frac{\phi_{k,1}+i\phi_{k,2}}{\sqrt{2}}$.
\begin{equation}
H=\frac{i}{2}\sum_{k=1}^N \epsilon_k \pa{\phi_{k,1}\phi_{k,2}-\phi_{k,2}\phi_{k,1}} = \sum_{k=1}^N \epsilon_k \pa{b_k^\dagger b_k - 1/2} = \sum_{k=1}^N \frac{\epsilon_k}{2}\tilde\sigma_k^z\,.
\label{eq:diag}
\end{equation}
This concludes the diagonalization of the Hamiltonian as a sum of $N$ \textit{fermionic modes}. Note that the full Hilbert space of dimension $2^N$ can be easily constructed by choosing the possible occupations of these $N$ modes. In particular, we can analytically read the energy of the ground state from Eq.~\eqref{eq:diag} as 
\begin{equation}
E_0=-\sum_{k=1}^N |\epsilon_k|/2\,,
\end{equation}
and use it to build the cost function of the optimization process.

\end{subappendices}
\end{cbunit}
\end{document}